\documentclass[]{emulateapj}
\newcommand{\acs}{$^{\prime\prime}$}
\newcommand{\kms}{km s$^{-1}$}

\newcommand{\Msol}{M$_\odot$}

\newcommand{\Zsol}{Z$_\odot$}

\slugcomment{Accepted by the Astrophysical Journal Letters}
\shortauthors{O'Neil \& Schinnerer}
\shorttitle{The First CO Map of an LSB Galaxy}
\begin{document}
\title{The First CO Map of a Low Surface Brightness Galaxy}
\date{31 March, 2003}
\author{K. O'Neil}
\affiliation{NRAO; PO Box 2; Green Bank, WV 24944 USA; {\it koneil@nrao.edu}}
%\footnote{Work done while at Arecibo Observatory}}
\author{E. Schinnerer}
\affiliation{Jansky Fellow; NRAO, P.O. Box 0, Socorro, NM 87801 USA; {\it eschinne@nrao.edu}}
%\footnote{Work done while at California Institute of Technology}}

\begin{abstract}
Using the Owens Valley Radio Observatory Millimeter-Wavelength Array (OVRO)
we have obtained the first CO map of a low surface brightness (LSB) galaxy.
The studied galaxy, UGC~01922, was chosen for these observations
because both of its previous CO detection with the IRAM 30m telescope
and its classification as a Malin 1 `cousin' -- an LSB galaxy
with M$_{HI}$ $\ge$ 10$^{10}$ M$_\odot$.  The OVRO map detected approximately 65\% 
of the CO(1$-$0) flux found earlier with the single dish measurements,
giving a detected gas mass equivalent to $\rm M_{H_2}\;=\;1.1\times10^9$ \Msol.
The integrated gas peak lies at the center of the galaxy and coincides with both the
optical and 1.4 GHz continuum emission peaks.
The molecular gas extends well beyond the OVRO
beam size ($\sim$4\acs\ or 3 kpc), covering $\sim$25\% of the optical bulge.
In all, perhaps the most remarkable aspect of this map is its 
unexceptional appearance.  Given that it took over ten years  to 
successfully detect molecular gas in any low surface brightness system,
it is surprising that the appearance and distribution of UGC 01922's CO
is similar to what would be expected for a high surface brightness
galaxy in the same morphological class.
\end{abstract}

\section{Introduction}

Low surface brightness (LSB) galaxies are typically defined as those galaxies
with observed central surface brightnesses of  $\mu_B(0)\;\geq$ 23.0 mag arcsec$^{-2}$.
Their global properties show LSB systems to be under-evolved when compared 
to their high surface brightness (HSB) counterparts, having high \ion{H}{1} 
gas mass-to-luminosity ratios, blue colors,  and typically low metallicities (Z $<$ 0.3 \Zsol)
\citep{deblok95,mcgaugh94,oneil00b,oneil97,oneil97b}.
It is likely that the slow stellar evolution in these galaxies is due, at least in part,
to the extreme environments within most LSB galaxies, such as gas densities
typically below the Toomre criteria for star formation \citep{oneil00a,deblok96} and extremely high
dark matter content even in the galaxies' inner radii \citep{deblok01}.
%
%It is generally accepted that star formation in galaxies
%proceeds within the dense molecular regions of the interstellar
%medium.  However, previous studies undertaken to look for molecular gas
%in the form of CO within LSB galaxies have found the
%CO content to be at least an order of magnitude lower than
%that of an HSB galaxy with similar global properties.   Assuming similar values for
%the H$_2$-to-CO conversion factor for LSB and HSB systems,
%this has led to belief in a lack of molecular cloud
%complexes in LSB systems, which in turn has raised the question 
%of whether the molecular ISM is involved at all with the star formation
%process of LSB galaxies \citep{oneil00a,deblok95}.

\begin{figure*}[ht]
\resizebox{\hsize}{!}{
\includegraphics{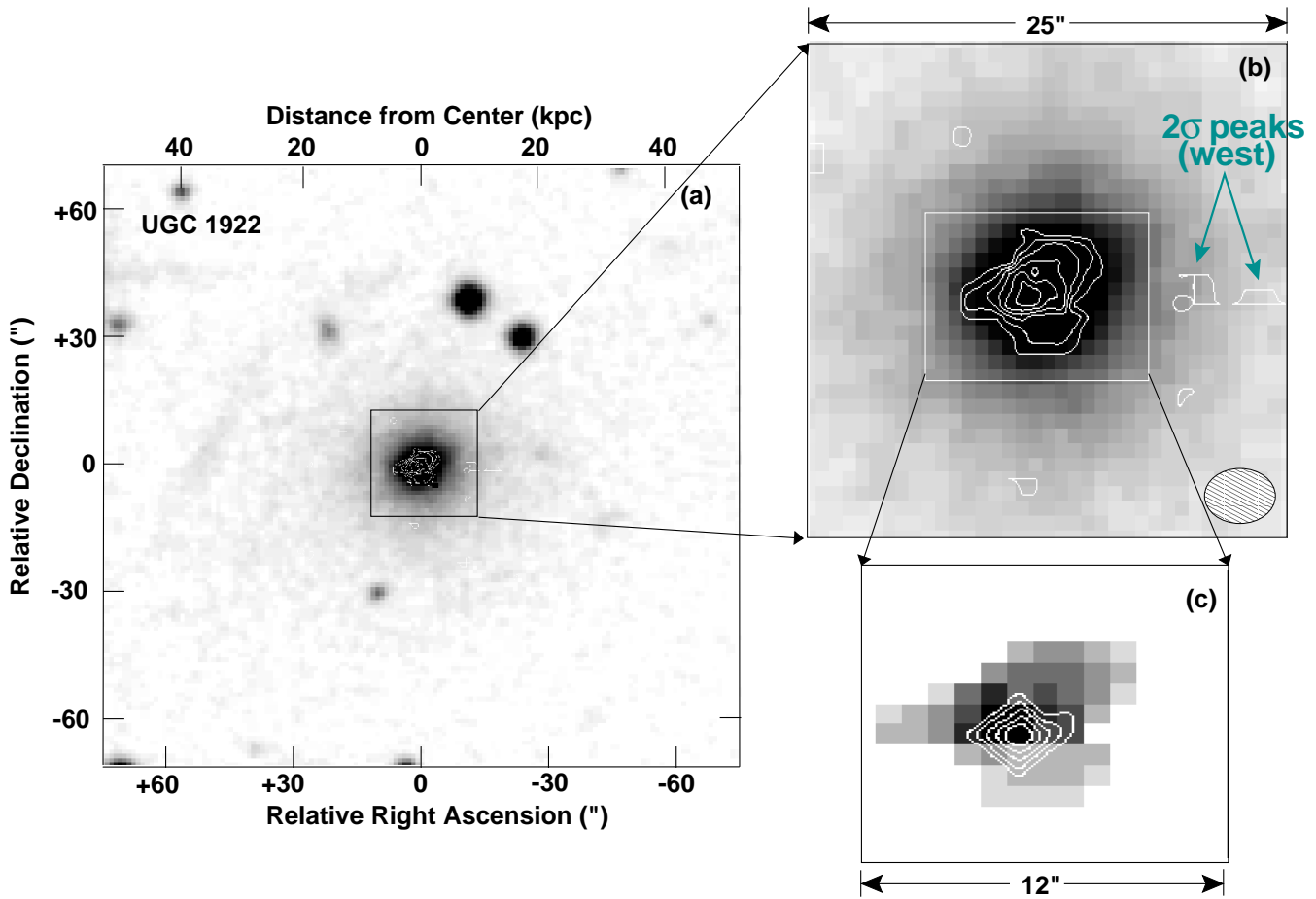}}
\caption{(a \& b) Grey-scale image of UGC~01922 from the DPOSS-II survey. 
Overlaid are contour lines showing the CO($1-0$) observations
integrated in velocity space. The contour lines lie at 0.48, 0.96, 1.44, 1.92, 2.40, and 2.88
Jy/beam \kms, with a 2$\sigma$ cut-off applied to the CO data.
Note that in 1a the spiral arms of UGC 01922 fill the image.
(c) At the bottom right, a gray-scale image of the OVRO map is shown with the 1400MHz
continuum contours overlaid.  The continuum contours lie at 6, 12, 18, 24, and 30 mJy.
Continuum data was obtained from the NRAO VLA Sky Survey (NVSS) \citep{condon98}.
\label{fig:overlay}}
\end{figure*}

After ten years of searching, molecular gas (in its CO emission) has finally
been detected in LSB galaxies \citep{oneil02,oneil00b}.
Taking advantage of these discoveries, we have used the Owens Valley
Radio Observatory Millimeter Array (OVRO) to obtain the
first CO map of an LSB system.  To date, it is {\it only} the 
massive LSB galaxies which have had CO detections \citep{oneil02}. 
As a result, the observed galaxy, UGC~01922, was chosen 
both because of its previous CO detection with the IRAM 30m telescope
and because it lies within the
class of massive LSB galaxies similar to the prototypical massive
LSB galaxy Malin 1.  This paper contains both a
description of the OVRO data as well as a discussion
as to the implications of our findings.

A Hubble constant of 70 km s$^{-1}$ Mpc$^{-1}$ was assumed throughout
this paper.  Additionally, a heliocentric velocity of
10,894 km s$^{-1}$ was assumed for UGC~01922, equating to a distance of 156 Mpc and
a scale of 0.75 kpc/\acs\ (assuming an infall velocity to the Virgo cluster
of 300 \kms).

\begin{figure}[ht]
\resizebox{\hsize}{!}{
\includegraphics{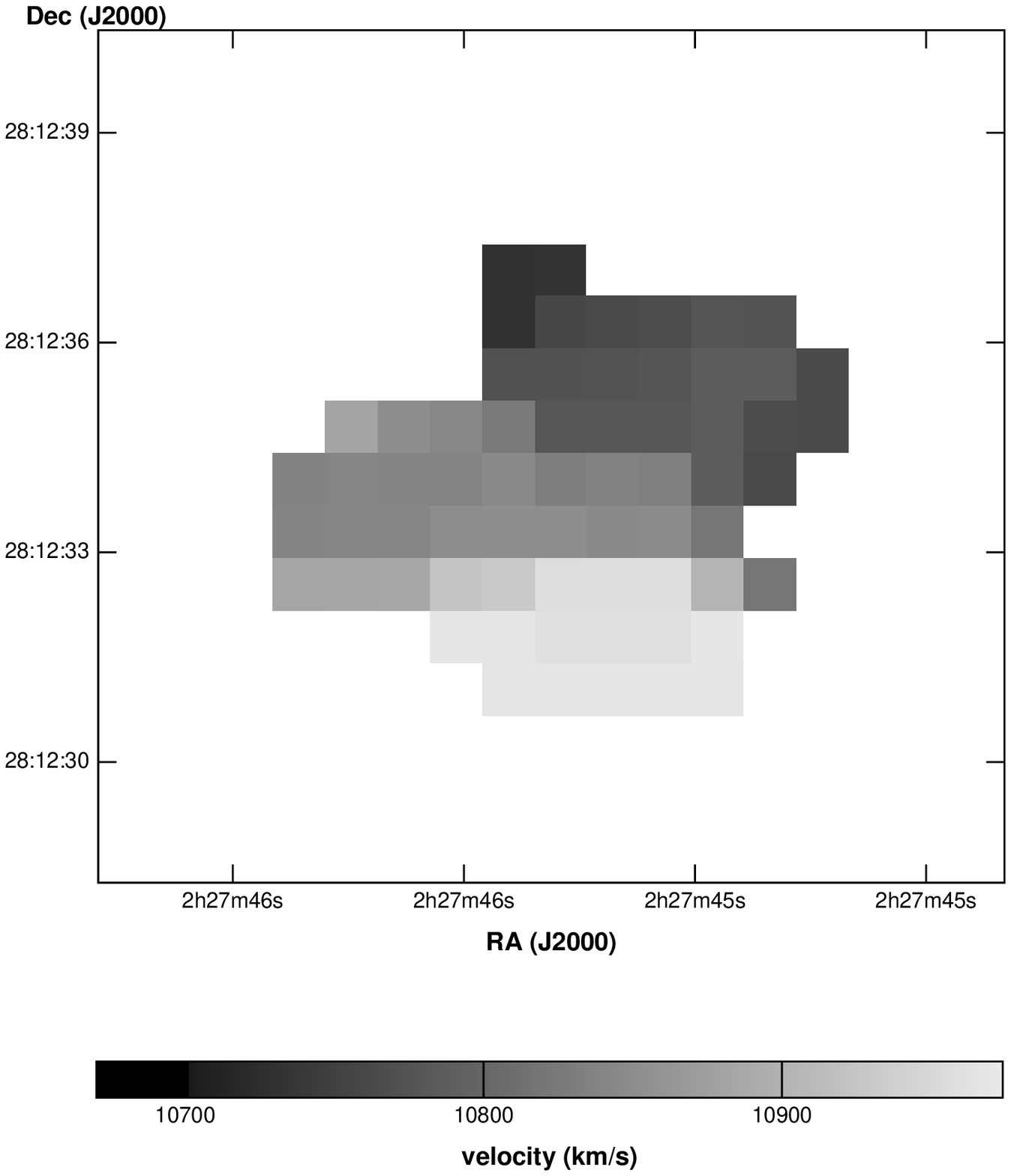}}
\caption{Iso-velocity map for UGC 10922 from the OVRO data.
\label{fig:mom1}}
\end{figure}

\section{Observations}

UGC~01922 was observed in its CO(1$-$0) line in 2002~March and April
using the six-element Owens Valley Radio Observatory (OVRO) millimeter
interferometer in its E and C configurations. The resulting baselines
range from 20m to 120m providing a spatial resolution of 4.5\acs\ $\times$ 3.3\acs\ 
(PA 88$^\circ$) using natural weighting within the primary beam of 65\acs\
at the observed frequency.
The quasars 3C84, 3C111, and 3C345 served as passband calibrators
and J0237+288 was observed every 20 minutes for phase calibration. The
average single sideband temperature was around 400\,K at the observed
line frequency of 111.23\,GHz.  The resulting noise per 30 \kms\
channel is $\sim$4 mJy beam$^{-1}$ in the combined data of 3
tracks.  The data were calibrated, mapped, and analyzed using the
software packages MMA \citep{scoville93}, MIRIAD \citep{sault95},
and GIPSY \citep{hulst92}, respectively.  The
intensity and iso-velocity maps contain all emission above the $2\sigma$ clipping level in
the velocity interval between $-$255 \kms\ and +165 \kms\
relative to the observed line velocity of v$_{LSR}$=10896 \kms\
(Figure~\ref{fig:overlay} \& Figure~\ref{fig:mom1}).

\begin{deluxetable}{llll}
\tablewidth{0pt}
\tablecolumns{4}
\scriptsize
\tablecaption{Properties of UGC 01922 \label{tab:props}}
\tablehead{}
\startdata
{$\bf RA$}....\dotfill 02:27:46.01$^1$ &
{$\bf Dec$}....\dotfill 28:12:30.3$^1$\\
{$\bf M_T^B$}\dotfill -19.8$^3$ &
{$\bf L_B$}\dotfill 6.6$\times\;10^{9}$ \\
{$\bf M_{HI}$}\dotfill $1.0\times\;10^{10}\;2$ & 
{$\bf M_{dyn}$}\dotfill 2.9$\times\;10^{12}\;2$\\ 
{$\bf R_{eff}^{bulge}$}\dotfill 18$^4$  & 
{$\bf R_T^{disk}$}\dotfill $\ge$59$^4$\\ 
{$\bf V_{HEL}$}\dotfill 10885$^2$& \\
\enddata
\tablerefs{\footnotesize $^1$\citet{oneil02}; $^2$\citet{giov85}; $^3$NASA Extragalactic
Database; $^4$Obtained from fitting isophotes to the Digital Sky Survey image of UGC 01922;
$^5$\citet{nilson73}; $^6$Infrared Science Archive (IRSA) Two Micron All Sky Survey (2MASS)
data; $^7$\citet{condon98}. }
\end{deluxetable}

\section{Known Properties of UGC 01922}

Classified as an Sb-Sd galaxy \citep{giov85},
UGC~01922 consists of a bright central core (bulge) surrounded by a very faint LSB
disk (Figure~\ref{fig:overlay}).  From the Digitized Palomar Sky Survey II
(DPOSS-II)\footnote{The Digitized Sky Surveys were produced
at the Space Telescope Science Institute under
U.S. Government grant NAG W-2166. The images of these surveys are based on
photographic data obtained using the Oschin Schmidt Telescope on Palomar Mountain and
the UK Schmidt Telescope. The plates were processed into the present compressed digital
form with the permission of these institutions.} image of UGC~01922, we find
the bulge effective
radius to be approximately 19 kpc (25\acs), while the LSB disk extends 
{\it at least} to a radius of 60 kpc (80\acs) and it may extend
considerably farther.  (The DPOSS-II images lack the surface brightness
sensitivity to determine the full extent of the galaxy's LSB disk.)
In contrast with the common belief that LSB galaxies are pure disk systems,
LSB galaxies like UGC~01922 with \ion{H}{1} masses in the range 
$\rm M_{HI} \ge 10^{10}$ \Msol\ are often found to have prominent
bulges \citep{knezek99}.  In fact, the overall appearance of
UGC~01922 is remarkably reminiscent of the morphology of Malin I,
the archetypal massive LSB system \citep{impey89}.
The previously measured properties of UGC 01922 are summarized in
Table~\ref{tab:props} and a comparison between UGC 01922's properties
and that of other galaxies is given in Table~\ref{tab:compare}.

In addition to having a strong presence in the optical images
of UGC~01922, the galaxy's central bulge shows up both as an
extended source in all bands of the 2MASS\footnote{The Two Micron All Sky Survey (2MASS)
is a joint project of the University of Massachusetts and the Infrared
Processing and Analysis Center/California Institute of Technology,
funded by the National Aeronautics and Space Administration and the
National Science Foundation.} survey data and as a strong
1400 MHz continuum emitter in the NRAO VLA Sky Survey (NVSS) \citep{condon98}.
The 2MASS data show emission in the central 49\acs\ (37 kpc),
roughly the same size as the central bulge as defined by the DPOSS-II 
image, with colors of $J-H\;=\;0.80\pm0.03,\;H-K_s\;=\;0.33\pm0.03$, and
$J$ = 11.60$\pm$0.02 (similar to the colors found in the inner regions of 
most HSB disk galaxies -- Hunt, et al. 1997).  No 2MASS emission is seen outside the galaxy's 
optically defined bulge.  Similarly, the NVSS emission extends across
the size of the optical bulge, having a deconvolved diameter (from the NVSS catalog)
of 40\acs\ (30 kpc) and a total flux of 36.5 mJy.  Both the 2MASS and NVSS catalogs
list a position angle of the emission as 68$^\circ$ (W of N).

Recent optical spectra taken across the nucleus of UGC~01922 
show \ion{N}{2}/H$\alpha$ $\sim$1, with 
accompanying strong \ion{S}{2} [$\lambda\lambda$ 6717,6731]
lines and a visible \ion{O}{1} [$\lambda$ 6300], all of which are
indicators of AGN or LINER emission within a galaxy.
(The optical spectra is shown in Figures 3 \& 4 of O'Neil, Schinnerer, \& Hofner 2003.)
This is not exceptionally surprising.  Schombert (1998) found that approximately
50\% of the massive LSB systems he studied have AGN/LINER nuclei,
while \citet{impey01} found $\sim$20\% of the massive LSB galaxies in
their study have LINER/AGN characteristics.  It is interesting to note that 
the underlying shape of UGC 01922's optical spectra contains a
strong 'bump' at 4700\AA\ similar to the
\ion{Fe}{3} bump found in late-type Type Ia supernovae \citep{r-l95, mazzali98}. 
This spectral shape is likely the result
of the supernova SN1989S which exploded within the galaxy
more than 10 years ago \citep{mueller89}, and should not have any effect
on the measurement of the relevant galaxy emission lines.

In all, UGC 01922 appears to be a fairly `typical' massive LSB galaxy with
AGN/LINER characteristics and a bulge which is similar to those found in many
HSB spiral galaxies.

\begin{deluxetable}{lccccc}
\scriptsize
\tablewidth{0pt}
\tablecaption{Comparison of the Mass-to-Luminosity Ratio of UGC 01922 with Other Galaxies \label{tab:compare}}
\tablehead{ \colhead{Property} & \colhead{UGC 01922} & \colhead{$\rm \langle LSB \rangle^1$} & 
\colhead{$\rm \langle HSB \rangle^2$} & \colhead{Malin 1$^3$} &\colhead{LSBC F568-6$^3$} }
\startdata
$\rm M_{HI}/L_B$ & 1.6* & 1.0 & 0.3 & 0.7 & 3.0\\
$\rm M_{dyn}/L_B$ & 36* & 11 & 0.05 & 24 & 28\\
\enddata
\tablecomments{\footnotesize *Both $\rm M_{HI}/L_B$ and $\rm M_{dyn}/L_B$ may be high 
(in part) due to the LSB nature of UGC 01922's disk resulting in an underestimate of L$_B$.}
\tablerefs{\footnotesize $^1$ \citet{deblok96}; $^2$\citet{broeils92}; $^3$\citet{bothun90}}
\end{deluxetable}

\section{Molecular Gas in UGC 01922}

\begin{figure}
\resizebox{\hsize}{!}{
\epsffile{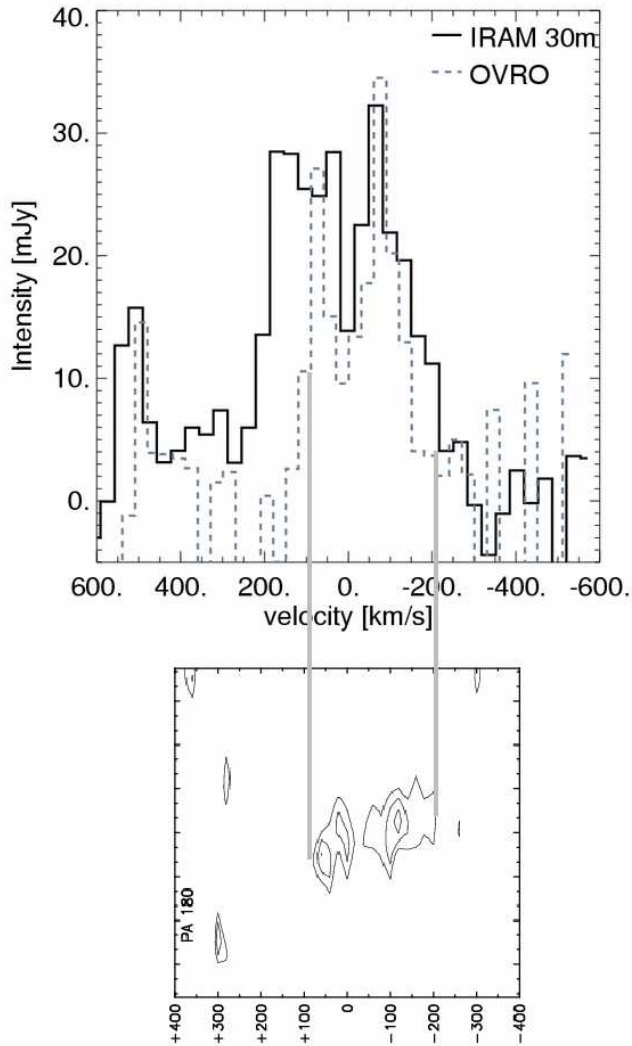}}
\caption{Plots of the CO velocity distribution.  At top, the black line is the IRAM
30m spectrum, from \citet{oneil02}, smoothed to 30 \kms\ resolution and the gray dashed line
is the integrated OVRO spectrum of the emission.   
At bottom is a position-velocity diagram of the OVRO data along a PA of 180$^\circ$
which shows the offset between the red and blue channels.   The y-axis is the relative
spatial offset, ranging from $-10^{\prime\prime}$ to $10^{\prime\prime}$.
The OVRO data is also binned to 30 km s$^{-1}$ resolution, and the
contours lie at 2$\sigma$, 3$\sigma$, \& 4$\sigma$ (1$\sigma$ = 4mJy/beam).
\label{fig:vel}} 
\end{figure}

Figure~\ref{fig:overlay} shows the integrated CO(1$-$0) emission for our
OVRO observations.  The rms noise of the (summed) intensity
map is  0.39 Jy/beam km s$^{-1}$, while the peak of the CO emission
lies at the 2.94 Jy/beam km s$^{-1}$
level, making our observations a clean 7$\sigma$ detection.
The total flux detected in the central 30\acs\ of the map is 3.6 Jy \kms,
or approximately 55\% of the total flux as found with the IRAM 30m telescope
(6.6 Jy km s$^{-1}$ -- O'Neil, Schinnerer, \& Hofner 2003). 
Including the tentative 2$\sigma$ peak seen at the western
edge of the map adds an additional 0.8 Jy \kms, bringing the total
flux seen with the OVRO map to $\sim$65\% of the IRAM flux.
As can be seen
in Figure~\ref{fig:vel}, the majority of the unrecovered flux appears
to lie in the wings of the line profile, and so it is likely to have too
low signal-to-noise to be reliably detected in the OVRO data.
The detected mass is equivalent to $\rm M_{H_2}\;=\;1.1\times10^9$ \Msol,
assuming a standard CO-to-H$_2$ conversion factor of
N(H$_2$)/$\int{T(CO)dv}$=3.6$\times$10$^{20}$~cm$^{-2}$~(K~\kms)$^{-1}$.
The central surface density, measured with one beam, is 6.3$\times$10$^{21}$ cm$^{-2}$,
or $\sim$50 \Msol\ pc$^{-2}$ using the same CO-to-H$_2$ conversion factor.

Examining Figure~\ref{fig:overlay} more closely we can see that
all the detected CO emission in UGC~01922 is contained within the
galaxy's bright central bulge, with the CO radius being approximately
25\% the bulge effective radius.  Figure~\ref{fig:overlay} also shows 
the integrated CO emission to be centered on the 
optical center of the galaxy, and coincident with both the 1400MHz continuum
emission found by the NVSS and the 2MASS near-IR emission.  Looking at
a position-velocity diagram along the kinematic major axis of the CO gas shows a 
spatial offset between the integrated peak emission of the red
and blue channels of 2 -- 3\acs\ ($\sim$1.5 -- 2.3 kpc -- Figure~\ref{fig:vel}),
indicating the CO gas lies in a ring around the galaxy's core.
As we do not have the resolution available at the other
wavelengths (1400 MHz, \ion{H}{1}, etc.), we cannot say whether 
this ring-like distribution is unique to the CO distribution or if it is also present
at other wavelengths.

The largest velocity gradient of the CO emission is along the north-south axis
which is not along the apparent elongation of the CO emission (Figure~\ref{fig:mom1}).
The uncorrected rotational velocity is 100 \kms, assuming the rotational
velocity is equal to half the difference between the maximum and minimum velocities 
(Figure~\ref{fig:mom1}).   
Assuming an inclination of 40$^\circ$ \citep{nilson73},
at a radius of 1.5 kpc (the edge of the centralized molecular gas in the OVRO map)
this gives an enclosed dynamical mass of 8.4$\times$10$^{9}$\Msol.
%or 1/20 the dynamical mass found by the \ion{H}{1} line of \citet{giov85}.
Using the above numbers, the detected molecular gas mass-to-dynamical mass ratio
for UGC~01922 is 0.13 for the central 3 kpc, a value similar to that typically 
found in HSB systems (e.g. \citet{sakamoto99} found $\langle M_{H_2}/M_{dyn} \rangle$ = 0.1
for the inner 1 kpc of 20 CO-luminous nearby spiral galaxies).

\section{Discussion}

CO is often considered to be a reliable tracer of the molecular gas within
a galaxy.  As molecular gas and molecular cloud complexes are directly associated
with star formation within galaxies, CO can also be considered a tracer of 
a galaxy's star formation potential.  Consequently, our understanding of the star
formation processes within LSB galaxies can be considerably increased by
knowing the general distribution of CO gas within these systems.  

The properties of UGC~01922's central bulge, including the strong
near-IR and radio continuum emission and the characteristic AGN/LINER 
spectra, indicate that unlike the disks of most LSB galaxies,
the density of baryons within the central few kpc of UGC~01922
must be quite high.  This is in contrast with both the LSB disk of UGC~01922 and the gas
density of most LSB galaxies, which is often below the Toomre
cut-off for star formation ($\rm \sim10^{21}\;cm^{-2}$)
\citep{oneil00a,deblok98,pickering97,vanzee97}.  

\citet{oneil02} compared all single dish observations of LSB galaxies with the
CO and H$_2$ properties of a representative HSB galaxy sample.  They
conclude that it is primarily the low baryonic density inherent in LSB disks
which inhibits high CO production.  Consequently, CO will only readily
be found in LSB galaxies which contain regions of high baryonic density.
UGC~01922, with its low density LSB disk surrounding a bright, high density nucleus,
is an excellent test case for the theory laid out in \citet{oneil02}.
In agreement with their theory, the only CO seen within UGC~01922 is in the                
higher density bulge, with no CO seen in the galaxy's low density disk.
The two exceptions to this idea -- Malin 1 and UGC 06968 --  are massive
LSB galaxies with properties similar to those of UGC 01922, yet which do
not have CO detections to fairly low limits.  In both of these
cases it is likely that the gas and dust is quite cold, which would lower the 
galaxies' CO-to-M$_{H_2}$ conversion factor.  A detailed discussion of 
this is given in \citet{oneil02}.

As mentioned in the last section, 
the maximum observed velocity reached by the CO gas is about 200 \kms.  A spatial
offset of approximately 2 kpc is found between the red and blue sides.
An offset of this size is seen
for circumnuclear star forming rings within galaxies (i.e. Regan, Sheth,
\& Vogel 1999; Planesas, Colina, \& P\'erez-Olea 1997). 
\citet{combes91} shows that circumnuclear star forming rings
are expected to form as a result of the increased gas density associated
with stellar bars.  The ring might host \ion{H}{2} regions
(regions of active star formation) which can change the radiation
field and metallicity of the area, resulting in our CO emission detection.
However, no evidence of a large-scale bar can be found in UGC~01922,
in either the DPOSS-II image (Figure~\ref{fig:overlay}) or a recently obtained
R-band image using the Lowell observatory's 1.8m telescope (O'Neil, Oey, \& Bothun, in prep.).

Within the HSB galaxy realm, a difference in the distribution of molecular gas with
galaxy type is often seen (e.g. Casoli, et al. 1998).  In particular, the earlier-type
disk galaxies typically have molecular gas concentrations in their central bulge
while the molecular gas of the later-type galaxies is often spread throughout the
galaxies' disks.  If one assumes that previous non-detections of molecular gas
in  LSB galaxies which do not contain any central bulge are due to the gas being 
both low in column density and distributed throughout the disk, it seems likely that
LSB galaxies have a similar CO distribution/morphology trend as the HSB galaxies.
The difference is simply in the density and possibly the total amount of molecular
gas, with less gas being detected within the environment of low surface brightness
galaxies \citep{oneil02}.  As a result, it seems as if the CO now being detected 
within LSB galaxies has the distribution and appearance which would be expected
for any galaxy, LSB or HSB, of its morphological type.

\acknowledgements{Many thanks to L. Ho for help in interpreting
the optical spectra of UGC~01922.}


\begin{thebibliography}{}

\bibitem[de Blok, McGaugh, \& Ruben(2001)]{deblok01}
de Blok, W.J.G., McGaugh, S., \& Ruben, V. 2001 AJ 122, 2381

\bibitem[de Blok \& van der Hulst(1998)] {deblok98}
de Blok, W.J.G. \& van der Hulst, J.M. 1998 A\&A 336 49

\bibitem[de Blok, McGaugh, \& van der Hulst(1996)]{deblok96}
de Blok, W.J.G., McGaugh, S., \& van der Hulst, J.M. 1996 MNRAS 283, 18

\bibitem[de Blok, van der Hulst, \& Bothun(1995)] {deblok95}
de Blok, W.J.G., van der Hulst, J.M., \& Bothun, G.D. 1995 MNRAS 274, 235

\bibitem[Bothun, et al.(1990)]{bothun90}
Bothun, G.D., Schombert, J., Impey, C. \& Schneider, S. 1990 ApJ 360, 427

\bibitem[Broeils(1992)]{broeils92}
Broeils, A. 1992 Ph.D. thesis (Rijksuniversiteit: Groningen)

\bibitem[Combes(1991)] {combes91} {Combes, F. 1991 {\it Dynamics of
Galaxies and Their Molecular Cloud Distribution}, ed. F. Combes \& F.
Casoli, p. 225, Kluwer}

\bibitem[Casoli, et al.(1998)]{casoli98}
Casoli, F., et.al 1998 A\&A 331, 451
% Sauty, S., Gerin, M., Boselli, A., Fouque, P., Braine, J.,
%Gavazzi, G., Lequeux, J., \&  Dickey, J 1998 A\&A 331, 451

\bibitem[Condon, et al.(1998)] {condon98}
Condon J. J., Cotton, W. D., Greisen, E. W., Yin, Q. F., Perley, R. A.,
Taylor, G. B., \& Broderick, J.  1998, AJ, 115, 1693

\bibitem[Giovanelli \& Haynes(1985)]{giov85}
Giovanelli, R. \& Haynes, M. 1985 AJ 90, 2445

\bibitem[Hunt, et al.(1997)]{hunt97}
Hunt, L. K., Malkan, M. A., Salvati, M., Palazzi, E.,
Mandolesi, N., \& Wade, R. 1997 ApJS 108, 229

\bibitem[Impey, Burkholder, \& Sprayberry(2001)]{impey01}
Impey, C., Burkholder, V., \& Sprayberry, D. 2001 AJ 122, 2341

\bibitem[Impey \& Bothun(1989)]{impey89}
Impey, C., \& Bothun, G. 1989 ApJ 341, 89

\bibitem[Kenezk(1999)]{knezek99}
Knezek, P. 1999 in {\it The Low Surface Brightness Universe, IAU Colloquium 171}
J.I. Davies, C. Impey, \& S. Phillipps, eds. (PASP:San Francisco)

\bibitem[Mazzali, et al.(1998)]{mazzali98}
Mazzali, P. A., Cappellaro, E., Danziger, I. J., Turatto, M., \& Benetti, S.
1998 ApJ 499, 49

\bibitem[McGaugh(1994)] {mcgaugh94}
McGaugh, S. 1994 ApJ 426, 135

%\bibitem[Mihos, Spaans, \& McGaugh(1999)]{mihos99}
%Mihos, C., Spaans, M., \& McGaugh, S.S. 1999 ApJ 515, 89

\bibitem[Mueller, Gunn, \& Oke(1989)]{mueller89} 
Mueller J., Gunn J., \& Oke, J. 1989 IAU Circulars 4888, 1 

\bibitem[Nilson(1973)]{nilson73}
Nilson, P. 1973 {\it Uppsala General Catalogue of Galaxies} (Uppsala: Astronomiska Observatorium)

\bibitem[O'Neil, Schinnerer, \& Hofner(2003)]{oneil02}
O'Neil, K., Schinnerer, E., \& Hofner, P. 2003, ApJ, accepted

\bibitem[O'Neil, Verheijen, \& McGaugh(2000)]{oneil00a}
O'Neil, K., Verheijen, M., \& McGaugh, S. 2000 AJ 119, 2194

\bibitem[O'Neil,  Hofner, \& Schinnerer(2000)]{oneil00b}
O'Neil, K., Hofner, P., \& Schinnerer, E. 2000 ApJ 545, L99

%\bibitem[O'Neil, Bothun, \& Schombert(2000)]{oneil00c}
%O'Neil, K., Bothun, G.D., \& Schombert, J. 2000 AJ 119, 136

\bibitem[O'Neil, et al.(1997)]{oneil97b}
O'Neil, K., Bothun, G.D., Schombert, J., Cornell, M., \& Impey, C. 1997 AJ 114, 2448

\bibitem[O'Neil, Bothun, \& Cornell(1997)]{oneil97}
O'Neil, K., Bothun, G.D., \& Cornell, M. 1997 AJ 113, 1212

\bibitem[Pickering, et al.(1997)]{pickering97}
Pickering, T. E.; Impey, C. D.; van Gorkom, J. H.; Bothun, G. D.
1997 AJ 114, 1858

\bibitem[Planesas, Colina, \& P\'erez-Olea(1997)]{planesas97}
Planesas, P., Colina, L., \& P\'erez-Olea, D. 1997 A\&A 325, 81

\bibitem[Regan, Sheth, \& Vogel(1999)]{regan99}
Regan, M., Sheth, K., \& Vogel, S. 1999 ApJ 526, 97

\bibitem[Ruiz-Lapuente, et al.(1995)]{r-l95}
Ruiz-Lapuente, P., Kirshner, R. P., Phillips, M. M., Challis, P. M., Schmidt, B. P.,
Filippenko, A. V., \& Wheeler, J. C. 1995 ApJ 439, 60

\bibitem[Sakamoto, et al.(1999)]{sakamoto99}
Sakamoto, K., Okumura, S., Ishizuki, S., \& Scoville, N.  1999 ApJ 525, 691

\bibitem[Sault, Teuben, \& Wright(1995)]{sault95}
Sault, R.J., Teuben, P. J., \& Wright, M. C. H. 1995
{\it Astronomical Data Analysis Software and Systems, Volume IV} (PASP:San Francisco)

\bibitem[Schombert(1998)]{schom98}
Schombert, J. 1998 AJ  116, 1650

\bibitem[Scoville et al.(1993)]{scoville93}
Scoville, N., et. al 1993 PASP 105, 1482

\bibitem[van der Hulst, et al.(1992)]{hulst92}
van der Hulst J. M., Terlou, J. P., Begeman, K. G., Zwitser, W., \&  Roelfsema, P. R. 1992
{\it Astronomical Data Analysis Software and Systems, Volume I} (PASP:San Francisco)

\bibitem[van Zee, et al.(1997)]{vanzee97}
van Zee, L., Haynes, M., Salzer, J., Broeils, A. 1997 AJ 113, 1618

%\bibitem[Wilson(1995)]{wilson95}
%Wilson, C. 1995 ApJ 391, 144
%
%\bibitem[Young \& Knezek(1989)]{yk89}
%Young, J. \& Knezek, P. 1989 ApJ 347, L55

\end{thebibliography}
\end{document}